# Gate tunable third-order nonlinear optical response of massless Dirac fermions in graphene

Tao Jiang, [1†] Di Huang, [1†] Jinluo Cheng,[2] Xiaodong Fan,[3,4] Zhihong Zhang,[5] Yuwei Shan,[1] Yangfan Yi,[1] Yunyun Dai,[1] Lei Shi,[1,6] Kaihui Liu,[5] Changgan Zeng,[3,4] Jian Zi,[1,6] J.E. Sipe,[7] Yuen-Ron Shen,[1,8] Wei-Tao Liu,[1,6*] Shiwei Wu[1,6*]

[1] State Key Laboratory of Surface Physics, Key Laboratory of Micro and Nano Photonic Structures (MOE), and Department of Physics, Fudan University, Shanghai 200433, China

[2] Changchun Institute of Optics, Fine Mechanics and Physics, Chinese Academy of Sciences, Changchun, Jilin 130033, China

[3] International Center for Quantum Design of Functional Materials, Hefei National Laboratory for Physical Sciences at the Microscale, and Synergetic Innovation Center of Quantum Information and Quantum Physics, University of Science and Technology of China, Hefei, Anhui 230026, China

[4] CAS Key Laboratory of Strongly-Coupled Quantum Matter Physics, and Department of Physics, University of Science and Technology of China, Hefei, Anhui 230026, China

[5] State Key Laboratory for Mesoscopic Physics and School of Physics, Peking University, Beijing 100871, China

[6] Collaborative Innovation Center of Advanced Microstructures, Nanjing 210093, China

[7] Department of Physics, University of Toronto, Toronto, Ontario, Canada

[8] Physics Department, University of California, Berkeley, CA 94720, USA

† T. Jiang and D. Huang equally contributed to this work.

* Corresponding email: swwu@fudan.edu.cn, wtliu@fudan.edu.cn

**Abstract**

Materials with massless Dirac fermions can possess exceptionally strong and widely tunable optical nonlinearities. Experiments on graphene monolayer have indeed found very large third-order nonlinear responses, but the reported variation of the nonlinear optical coefficient by orders of magnitude is not yet understood. A large part of the difficulty is the lack of information on how doping or chemical potential affects the different nonlinear optical processes. Here we report the first experimental study, in corroboration with theory, on third harmonic generation (THG) and four-wave mixing (FWM) in graphene that has its chemical potential tuned by ion-gel gating. THG was seen to have enhanced by ~30 times when pristine graphene was heavily doped, while difference-frequency FWM appeared just the opposite. The latter was found to have a strong divergence toward degenerate FWM in undoped graphene, leading to a giant third-order nonlinearity. These truly amazing characteristics of graphene come from the possibility to gate-control the chemical potential, which selectively switches on and off one- and multi-photon resonant transitions that coherently contribute to the optical nonlinearity, and therefore can be utilized to develop graphene-based nonlinear optoelectronic devices.

Graphene exhibits extraordinarily strong coupling to light owing to its unique linear and gapless two-dimensional band structure that hosts massless Dirac fermions[1,2]. The wide band linearity results in a unique spectral response ranging from terahertz/infrared to the visible and UV. The corresponding optical absorbance is a constant of a universal value of 2.3% for a suspended graphene monolayer[3,4]. Gate tuning of the carrier density, and hence the chemical potential (or Fermi level), modifies both intraband and interband transitions in graphene, and allows control of its optical properties in selected spectral regime, leading to many promising applications in optoelectronics and photonics[1,2,5-9].

The very strong linear response in such regimes suggests that the nonlinear optical response of graphene could also be exceptionally strong and promising for optoelectronic applications[10,11]. Since second-order nonlinearity in graphene is electric-dipole forbidden because of inversion symmetry, the third-order nonlinear optical response becomes dominant. Indeed, third harmonic generation (THG)[12-15], four-wave mixing (FWM)[16,17], optical Kerr effect[18-21], self-phase modulation (SPM)[22,23], two-colour coherent optical injection of current[24], and even high harmonic generations (HHG)[25] can be readily observed in graphene. However, the reported values of the third-order nonlinear response coefficients appear to vary by more than 6 orders of magnitudes (see Table S1 in the SI). It is not yet clear how such a wide variation comes about despite differences between the nonlinear processes studied and the experimental conditions employed. A unified understanding of the nonlinear optical response of graphene is needed, and is crucial for future design of graphene-based nonlinear photonic devices[6,22]. It will also provide a salient platform for study of the third-order nonlinear optical response of massless Dirac fermions that exist in other novel materials such as topological insulators[26], Dirac and Weyl semimetals[27].

For a better understanding of the third-order optical response of graphene, we must know how it varies with input frequencies with respect to the chemical potential. Tuning the input frequencies or the chemical potential can move a third-order nonlinear process in and out of one-, two-, or three-photon resonances, and provide detailed information about the nonlinear process. Unfortunately, such experiments have not yet been reported, although they have been suggested in theoretical work[28-32].

Here we report the first experimental study of the third-order nonlinear optical response of ion-gel gated graphene. Our focus is on THG and FWM, but the extension to other third-order processes is straightforward. The ion-gel gating allowed us to controllably tune the chemical potential over a sufficiently large range such that one-, two-, and three-photon resonances could be selectively turned on or off for a given set of input frequencies. We found that THG in the heavily doped graphene could be much stronger (~30 times) than in undoped graphene, while difference-frequency FWM behaved just the opposite, and exhibited a strong divergence of nonlinearity toward the degenerate FWM in undoped graphene. Our experimental results matched well with the theoretical calculation following Ref. 28, 29. Thus this work provides a firm basis for comprehension of third-order nonlinear optical processes in graphene and graphene-like Dirac materials.

**Shift of Chemical Potential by Gate Tuning**

For gate tuning of the chemical potential, we adopted the ion-gel gating method using the field effect transistor structure with graphene supported by fused silica[9, 33, 34], as depicted in Fig. 1a. This device structure enabled us to measure the linear and nonlinear optical responses of graphene at room temperature and monitor *in situ* the chemical potential $\mu$ versus the gate voltage $V_g$. Figure 1b plots the graphene resistance as a function of $V_g$. At $V_g \approx 0.9$ V, the resistance is maximum, indicating that graphene is at the charge neutral point (CNP, $\mu = 0$). Away from the CNP, the resistance decreases and $\mu$ shifts to positive or negative value accordingly.

To extract the chemical potential $\mu$ as a function of gate voltage, we measured the transmittance spectra of the gated sample at normal incidence[35]. The spectra at different $V_g$ normalized against the one at $V_{CNP}$ are shown in Fig. 1c. As described in Fig. 1d, the interband transitions should be suppressed by Pauli blocking for optical frequency $\hbar\omega_0 < 2|\mu|$ at zero temperature, resulting in a step-like transmission spectrum. At finite temperature, the spectrum is thermally broadened into a shoulder-like one, as seen in Fig. 1c. We could use the Kubo formula at T = 300 K to fit each spectrum and deduce $|\mu|$ from the fitting (described in the SI)[35]. The deduced $|\mu|$ as a function of $V_g$ is plotted in Fig. 1b. The result agrees well with that (red curve in Fig. 1b) predicted from an ion-gel gated graphene device with a capacitance of 2.5 μF/cm$^2$ (see the SI). The uncertainty of $|\mu|$ so obtained was ±10 meV. Ion-gel gating permitted us to tune $|\mu|$ from 0 to ~0.9 eV, corresponding to a tuning of the carrier density of graphene from 0 to ~6×10$^{13}$ cm$^{-2}$.

**Experiment on Third Harmonic Generation (THG)**

The linearly polarized femtosecond laser at 1566 nm ($\hbar\omega_0 = 0.794$ eV) was used to excite THG of ion-gel gated graphene at normal incidence, as described in Methods. Two representative output spectra taken in the reflected direction at $\mu = 0$ and $\mu = -0.74$ eV are displayed in Fig. 2a. The former shows a THG peak at 2.381 eV superimposed on a broad background, which is absent in the latter. The broad background is known to be due to ultrafast photoluminescence arising from Auger-like scattering of one-photon excited carriers[36]. It disappears when $2|\mu|$ is larger than $\hbar\omega_0$ so that the one-photon excitation is Pauli-blocked. While the THG peak was readily observable at all $\mu$ (Fig. 2b), its intensity exhibits shoulder-like rises as $2|\mu|$ moves over $\hbar\omega_0$ and $2\hbar\omega_0$ and reaches a maximum strength of ~30 times that of $\mu = 0$, as seen from the curves plotted in Fig. 2c for four different input wavelengths, 1300 nm (0.956 eV), 1400 nm (0.888 eV), 1566 nm (0.794 eV) and 1650 nm (0.753 eV). As will be explained more clearly later, these shoulder-like features arise from stepwise switching off of resonant transitions in graphene when $|\mu|$ increases: one-photon, two-photon, and three-photon resonant transitions are switched off successively when $2|\mu|$ becomes larger than $\hbar\omega_0$, $2\hbar\omega_0$, and $3\hbar\omega_0$ (Unfortunately, the last step could not be reached in our experiment.). Note that without graphene on the substrate, THG from the ion-gel gated fused silica was not observable.

The dependence of THG on input/output polarization is governed by the $D_{6h}$ structural symmetry of graphene. We found that if the normally incident input was linearly polarized and the analyser for the reflected THG was set at an angle $\theta$ with respect to the input polarization, the observed THG output was proportional to $\cos^2\theta$. Figure 2d presents two examples of THG versus $\theta$ taken at $\mu = -0.89$ eV with input polarizations along and perpendicular to the source-drain direction (Fig. 1a), respectively. In another measurement, we set the output analyser parallel to the input polarization and rotated them together with respect to the sample about its surface normal. The observed THG was isotropic, independent of the azimuthal rotation (Fig. 2e). Both results can be understood knowing that the third-order nonlinear susceptibility element, $\chi^{(3)}_{xxxx}$, of graphene is responsible for the THG (see the SI).

**Experiment on four-wave mixing (FWM)**

Four-wave mixing with two input frequencies $\omega_1$ and $\omega_2$ ($\omega_1 > \omega_2$) is a more general process than THG, but the effect of shifting $\mu$ to switch resonant transitions on and off is similar. Four FWM processes, described in Fig. 3a, are considered here: two sum-frequency mixings (SFM) with output at $2\omega_1 + \omega_2$ and $\omega_1 + 2\omega_2$, and two difference-frequency mixings (DFM) with output at $2\omega_1 - \omega_2$ and $2\omega_2 - \omega_1$. In our experiment, we chose $\hbar\omega_1 = 1.195$ eV (1040 nm) and $\hbar\omega_2 = 0.956$ eV (1300 nm), which generated SFM outputs at 3.346 eV (371 nm) and 3.107 eV (400 nm), and DFM outputs at 1.434 eV (867 nm) and 0.717 eV (1734 nm). The last DFM output was outside our spectral detection range. To study this process, we slightly shifted $\hbar\omega_2$ to 0.994 eV (1250 nm) to generate DFM ($2\omega_2 - \omega_1$) at 0.794 eV (1566 nm). The observed spectra taken at $\mu$

= 0 and $\mu = -0.73$ eV for the four mixing processes are displayed in Fig. 3b-d, showing the respective spectral peaks.

The SFM peaks are much stronger at $|\mu| = 0.73$ eV than at $\mu = 0$, but the DFM peaks show the opposite trend. The SFM processes are expected to be quite similar to THG, exhibiting a shoulder-like rise as $2|\mu|$ approaches $\hbar\omega_1$ and $\hbar\omega_2$ (Individual shoulders merge into one because of thermal broadening.). This is seen for the $\hbar(\omega_1 + 2\omega_2)$ process in Fig. 3e. The curve shows another rise as $2|\mu|$ approaches $2\hbar\omega_1$, $2\hbar\omega_2$, and $\hbar(\omega_1 + \omega_2)$. Unfortunately, the top of the rise cannot be seen because it was outside the tuning range of $|\mu|$. The DFM processes behave oppositely: at $\mu = 0$, the output is strong, but as $2|\mu|$ moves toward $\hbar\omega_1$, $\hbar\omega_2$, $\hbar(2\omega_1 - \omega_2)$ or $\hbar(2\omega_2 - \omega_1)$, it shows a step-like drop, as seen in Fig. 3e for the $2\omega_1 - \omega_2$ process and Fig. 3f for both DFM processes.

**Theoretical understanding and comparison with experiment**

To understand the observed $\mu$-dependences of THG and FWM in graphene in depth, we resort to the theory developed by Cheng et al.[28, 29]. The analytical expression of the third-order nonlinear susceptibility, $\chi^{(3)}$, generally has 8 terms for THG and 24 terms for the FWM processes studied here[37, 38]. In our case, $\chi^{(3)}$ of graphene is dominated by contributions from interband transitions; with the gapless, linearly dispersed bandstructure, each term in $\chi^{(3)}$ can only have a single resonance at either $\omega_i = 2v_F|k|$ or $|\omega_i \pm \omega_j| = 2v_F|k|$ or $|2\omega_i \pm \omega_j| = 2v_F|k|$ that provides the resonant enhancement. Here, $\omega_i$ and $\omega_j$ refer to the input frequencies, $v_F$ is the Fermi velocity in graphene, and $k$ is the electron wavevector in the first Brillouin zone. The above-mentioned resonances can be switched off by Pauli blocking if $2|\mu|$ becomes larger than $\hbar\omega_i$, $|\hbar\omega_i \pm \hbar\omega_j|$, and $|2\hbar\omega_i \pm \hbar\omega_j|$, respectively. It is expected that switch-off of resonances will introduce characteristic changes of $\chi^{(3)}$.

Mathematically, we can write a single resonant term in $\chi^{(3)}$ in the form of $\int \frac{A(k)\Delta n(k,\mu)}{\omega_I - 2v_F|k|} d^2k$, where $\omega_I$ is the input frequency or frequency combination on interband resonant transition and $\Delta n(k,\mu)$ is the difference of Fermi distributions of electrons between valence and conduction bands. The dependence of $\chi^{(3)}$ on $\mu$ is through $\Delta n(k,\mu)$, which has a derivative $\partial \Delta n(k,\mu)/\partial \mu \sim \delta(|\mu| - v_F|k|)$ at T ~ 0 K. We then have $\partial \chi^{(3)}(\mu)/\partial \mu \propto \frac{1}{2\mu - \hbar\omega_I}$. Since $\frac{1}{2\mu - \hbar\omega_I} = \mathcal{P}\frac{1}{2\mu - \hbar\omega_I} + i\pi\delta(2\mu - \hbar\omega_I)$ for $2\mu \sim \hbar\omega_I$, with $\mathcal{P}$ denoting the principal value, we find $\chi^{(3)} \propto [\ln|2\mu - \hbar\omega_I| + i\pi H(2\mu - \hbar\omega_I)]$ as $2\mu$ moves across $\hbar\omega_I$, where $H(2\mu - \hbar\omega_I)$ is the Heaviside step function, equal to 0 for $2|\mu| < \hbar\omega_I$ and 1 for $2|\mu| > \hbar\omega_I$, and $\ln|2\mu - \hbar\omega_I|$ exhibits a divergent peak at $|2\mu| = |\hbar\omega_I|$. Introduction of finite temperature and resonant damping effects will round up the step and smear the peak, making $|\chi^{(3)}|$ versus $\mu$ appear as shoulder-like rise seen in our experiment. The full mathematical derivation of $\chi^{(3)}$ for graphene, including contributions from both interband and intraband transitions, has been worked out by Cheng *et al.*[28], which is sketched in the SI. We have carried out calculation following their theory to compare with our experimental results.

Consider THG first, which is the simplest among all third-order processes. The analytic expression of Cheng *et al.* has a concise form of $\chi^{(3)}_{xxxx} \propto \left[-17G\left(\frac{\hbar\omega_0}{2|\mu|}\right) + \right.$

$64G\left(\frac{2\hbar\omega_0}{2|\mu|}\right) - 45G(\frac{3\hbar\omega_0}{2|\mu|})\right]$ with $G(x) = \ln\left|\frac{1+x}{1-x}\right| + i\pi H(|x|-1)$. The three G terms in the brackets describe switching off of one-, two- and three-photon resonant transitions as $2|\mu|$ moves over $\hbar\omega_0$, $2\hbar\omega_0$, and $3\hbar\omega_0$, as illustrated in Fig. 4a. Note that the sign of the first and third terms for one-photon and three-photon resonant transitions is opposite to that of the second term for two-photon resonant transitions. When $2|\mu| < \hbar\omega_0$, all three terms contribute to $\chi^{(3)}_{xxxx}$, but they nearly cancel each other, leaving $\chi^{(3)}_{xxxx}$ very small. With $\hbar\omega_0 < 2|\mu| < 2\hbar\omega_0$, one-photon resonant transitions are blocked and $G\left(\frac{\hbar\omega_0}{2|\mu|}\right) \sim 0$; imperfect cancellation of the $G\left(\frac{2\hbar\omega_0}{2|\mu|}\right)$ and $G\left(\frac{3\hbar\omega_0}{2|\mu|}\right)$ terms leads to a significant positive value of $\chi^{(3)}_{xxxx}$. With $2\hbar\omega_0 < 2|\mu| < 3\hbar\omega_0$, the value of $|\chi^{(3)}_{xxxx}|$ increases further as both one-photon and two-photon resonant transitions are blocked with $G\left(\frac{\hbar\omega_0}{2|\mu|}\right) \sim 0$ and $G\left(\frac{2\hbar\omega_0}{2|\mu|}\right) \sim 0$. Finally, for $2|\mu| > 3\hbar\omega_0$, all resonant transitions are blocked, leaving again a vanishingly small $\chi^{(3)}_{xxxx}$ from nonresonant contributions. The calculated $\mu$-dependence of $\chi^{(3)}_{xxxx}$ with $\hbar\omega_0 = 0.956$ eV is plotted in Fig. 4b. While it captures the essence of the THG response, the detailed shape of the curve is far from reality because resonant damping and finite temperature effects have been neglected. For better comparison with experiment, we include in the calculation the finite temperature effect on $\Delta n(\mu)$ (T = 300 K) and proper resonant damping factors ($\Gamma_e = 100|\mu|$ meV with $\mu$ in eV[39], and $\Gamma_i = 0.5$ meV for interband and intraband resonances, respectively). The calculated $|\chi^{(3)}_{xxxx}|$ versus $\mu$ is shown together with the data deduced from experiment in Fig. 4c. Both reveal the features mentioned earlier.

Similar discussion can be applied to FWM. The third-order susceptibility $\chi^{(3)}_{xxxx}$ for the two-colour SFM ($2\omega_1 + \omega_2$ and $\omega_1 + 2\omega_2$) increases with $|\mu|$ as for THG, but there are 5 resonant transitions for each SFM process, including: the one-photon transitions at $\hbar\omega_1$ and $\hbar\omega_2$, two-photon transitions at $2\hbar\omega_1$ (or $2\hbar\omega_2$) and $\hbar(\omega_1 + \omega_2)$, and three-photon transitions at $\hbar(2\omega_1 + \omega_2)$ (or $\hbar(\omega_1 + 2\omega_2)$). Figure 5a shows the calculated $\chi^{(3)}_{xxxx}$ versus $\mu$ for $\omega_1 + 2\omega_2$ SFM at zero temperature and without resonant damping. The characteristic features around the 5 specific values of $\mu$ are clearly seen. The expression of $\chi^{(3)}_{xxxx}$ for SFM is given by Eq. (3-4) in the SI. Again, the terms for two-photon transitions have opposite sign with respect to the terms for one- and three-photon transitions, leading to a much weaker $\left|\chi^{(3)}_{xxxx}\right|$ when $2|\mu| < \hbar\omega_1$ and $\hbar\omega_2$. Given the finite temperature (300 K) effect and resonance damping, resonances due to one-photon and two-photon transitions are greatly smeared, as shown in Fig. 5b. The theoretical simulation reasonably agrees with the experimental result plotted in Fig. 5b.

In sharp contrast to SFM, the DFM processes ($2\omega_1 - \omega_2$ and $2\omega_2 - \omega_1$) show opposite $\mu$-dependence with the output strongest at $\mu \sim 0$. The expression of $\chi^{(3)}_{xxxx}$ for DFM is the same as that for SFM except for a flip of sign on $\omega_1$ or $\omega_2$ (see the SI). Pauli blocking occurs at $2|\mu| > \hbar\omega_1$ and $\hbar\omega_2$ for one-photon transitions, $2|\mu| > 2\hbar\omega_1$ (or $2\hbar\omega_2$) and $\hbar(\omega_1 - \omega_2)$ for two-photon transitions, and $2|\mu| > \hbar(2\omega_1 - \omega_2)$ (or $\hbar(2\omega_2 - \omega_1)$) for three-photon transitions. The corresponding characteristic features can again be seen in the calculated $\chi^{(3)}_{xxxx}$ versus $\mu$ (Fig.5c for the $2\omega_1 - \omega_2$ DFM process). Note that the feature at $2|\mu| = \hbar|\omega_1 - \omega_2|$ is present, but is very weak and hardly visible in Fig. 5c, because it is described by a $G\left(\frac{\hbar|\omega_1-\omega_2|}{2|\mu|}\right)$ term with a coefficient proportional to $(\omega_1 - \omega_2)^3$. Increase or decrease at each step of the change depends on the sign of the frequency factor associated with each type of transitions. It is seen that for the $2\omega_1 - \omega_2$ DFM process, there are three terms in the equation for $\chi^{(3)}_{xxxx}$ that have the frequency factor $(\omega_1 - \omega_2)^2$ in the denominator. They contribute dominantly to $\chi^{(3)}_{xxxx}$ when $2|\mu| < \hbar\omega_2$, especially if $\omega_2$ is close to $\omega_1$, and yield a large step change when each term drops off at a specific value of $|\mu|$ because of Pauli blocking of the specific type of resonant transitions. The exceptionally large $\chi^{(3)}_{xxxx}$ for DFM is in strong contrast to the very weak $\chi^{(3)}_{xxxx}$ for SFM. Inclusion of the finite temperature effect and resonance damping in the calculation of $\chi^{(3)}_{xxxx}$ as a function of $\mu$ again smears out the peaks and spreads out the curve. The calculated curve of $|\chi^{(3)}_{xxxx}|$ versus $\mu$ agrees fairly well with the experimental result in Fig. 5d.

We note that as long as $2|\mu| < \hbar\omega_1$ or $\hbar\omega_2$, the $2\omega_1 - \omega_2$ and $2\omega_2 - \omega_1$ DFM would appear divergent through the frequency factor $(\omega_1 - \omega_2)^{-2}$ as $\omega_1$ approaches $\omega_2$ (Eq. (3-6) or (3-7) in the SI). One therefore expects that degenerate FWM including self-phase modulation would be extraordinarily strong in undoped graphene ($\mu = 0$). This was not noticed in the early pioneering work of Hendry et al.[16]. To experimentally verify such a behavior, we measured DFM of $2\omega_1 - \omega_2$ with $\mu$ close to zero, $\omega_1$ fixed at 1.195 eV (1040 nm), and $\omega_2$ tuned from 0.956 eV (1300 nm) to 1.11 eV (1120 nm). As shown in Fig. 5e, $\chi^{(3)}_{xxxx}$ for DFM increased by ~3 times as $\Delta\omega = \omega_1 - \omega_2$ decreased and agrees fairly well with the theoretical calculation. We expect a more rapid rise of DFM if DFM at smaller $\Delta\omega$ could be measured.

We adopted the scheme of Ref. 12 to measure the average output power of THG and FWM and estimated the value of $|\chi^{(3)}_{xxxx}|$ for the processes (See the SI). We found for THG at $\hbar\omega_0 = 0.956$ eV and $2|\mu| < \hbar\omega_0$, $|\chi^{(3)}_{xxxx}| = 1.9 \pm 0.3 \times 10^{-19}$ m$^2$/V$^2$. This value is very close to the theoretical value of $|\chi^{(3)}_{xxxx}| = 1.0 \times 10^{-19}$ m$^2$/V$^2$. In comparison, our value is about 2-3 orders of magnitude smaller than that of Kumar *et al.*[12], but consistent with the recent work of Woodward *et al.*[15], assuming $2|\mu| < \hbar\omega_0$ was satisfied in their experiments. For DFM, our experimental value of $|\chi^{(3)}_{xxxx}|$ is also close to the theoretical one as seen in Fig. 5e.

**Conclusions**

We have demonstrated that the third-order nonlinearity of graphene is exceptionally large and can be varied by orders of magnitude with the help of gate-controlled doping or shift of the chemical potential. The results can be understood from a unified theory on four-wave mixing in graphene. It is now possible to well predict the dependence of the third-order nonlinear responses of graphene on input frequencies and doping level. The understanding can be extended to other nonlinear optical processes in graphene, such as effective second-order processes[40-42], and even high-order harmonic generation[25]. Generally, the optical nonlinearity of linear-band materials with the chemical potential close to the Dirac or Weyl point tends to diverge in cases where input frequency combination approaches zero. The resulting giant nonlinearity of such materials, particularly graphene, can be of great use in future optoelectronic devices[6].

**Methods**

1. **Device fabrication**

Single crystalline[43, 44] or polycrystalline[45] graphene monolayers used in the experiment were grown by chemical vapour deposition (CVD) and transferred onto fused silica substrates. Source, drain and gate electrodes (50-nm Au and 5-nm Cr) were patterned through a dry stencil mask by electron beam deposition. All the electrodes were wire-bonded to a chip carrier for electrical control. Ion-gel gating was achieved by uniformly applying freshly prepared ion-gel solution onto the graphene devices, and further drying in a glove box filled with high purity argon gas. The ion-gel solution was prepared by dissolving 16.7 mg of Poly(styrene-b-ethylene oxide-b-styrene) (PS-PEO-PS) and 0.5 g of 1-Ethyl-3-methylimidazolium bis(trifluoromethylsulfonyl)imide ([EMIM][TFSI]) into 1.82 ml of dry dichloromethane. PS-PEO-PS, [EMIM][TFSI] and dry dichloromethane were purchased from J&K Scientific. Experimental results of THG and FWM from single crystalline and polycrystalline graphene, as well as on exfoliated monolayer, were found to be very much the same.

2. **Characterization and measurement**

The device characterization and experimental measurement were conducted in sample scanning optical microscopes that combined with femtosecond laser systems and an electrical transport setup. During the whole measurement, the graphene device was maintained in a dry nitrogen environment at room temperature. The charge neutral point of graphene was determined by its maximum resistance in response to the gate voltage as shown in Fig. 1a. A Fourier transform infrared (FTIR) spectrometer (VERTEX 70) was used to measure the transmittance spectra of gated graphene, from which the chemical potential was deduced as described in the main text and in the SI.

For THG measurements, a linearly polarized femtosecond laser beam (MaiTai HP and Inspire Auto, Spectra Physics) tunable from 345 to 2500 nm was focused and normally incident on graphene through a microscopic objective (100×, NA 0.95, Nikon), and the reflected THG signal was collected. The sample sitting on a nano-positioning stage enabled us to locate defect-free areas on the sample. A single-photon counting silicon avalanche photodetector (Perkin-Elmer) or a fiber-coupled spectrograph equipped with a liquid nitrogen cooled silicon charge-coupled device (Princeton Instruments) was used to detect the THG signal after proper filtering. The detailed optical arrangement is depicted in Fig. S1a of the SI. For measurement of the polarization-dependent azimuthal pattern of THG measurement (displayed in Figs. 2d and 2e), the transmitted THG geometry was adopted with the setup sketched in Fig. S1b of the SI.

For FWM measurements, a different femtosecond laser system (Insight Deepsee, Spectra Physics) was used, which could simultaneously produce two beams of different wavelengths at a repetition rate of 80 MHz, one tunable from 700 to 1300 nm and the other fixed at 1040 nm. The two beams were sent collinearly on the sample at normal incidence through a scanning optical microscope and the reflected FWM signal was detected. For the DFM signals in Fig. 3d and 3f, the spectra were recorded by a fiber-coupled spectrograph equipped with a liquid nitrogen cooled InGaAs array detector (PyLoN-IR, Princeton Instruments).

**Acknowledgements**

The work at Fudan University was supported by the National Basic Research Program of China (Grant Nos. 2014CB921601, 2016YFA0301002), National Natural Science Foundation of China (Grant No. 91421108, 11622429, 11374065), and the Science and Technology Commission of Shanghai Municipality (Grant No. 16JC1400401). Part of the sample fabrication was performed at Fudan Nano-fabrication Laboratory. K.H.L. is supported by National Natural Science Foundation of China (Grant No. 51522201). J.E.S. is supported by the Natural Sciences and Engineering Research Council of Canada. Y.R.S. acknowledges support from the Director, Office of Science, Office of Basic Energy Sciences, Materials Sciences and Engineering Division, U.S. Department of Energy (contract no. DE-AC03-76SF00098).

**Author contributions**

S.W.W. and W.-T.L. conceived and supervised the project. T.J., D.H., Y.W.S. and Y.F.Y. prepared the devices and performed the experiments, with assistance from Y.Y.D., L.S. and J.Z. on gate-dependent optical transmittance measurement. X.D.F., Z.H.Z., K.H.L. and C.G.Z. provided the CVD-grown graphene samples. T.J., D.H., J.L.C., J.E.S., Y.R.S., W.-T.L. and S.W.W. analysed the data. T.J., D.H., J.E.S., Y.R.S., W.-T.L. and S.W.W. wrote the paper with contributions from all authors.

**Figures and legends**

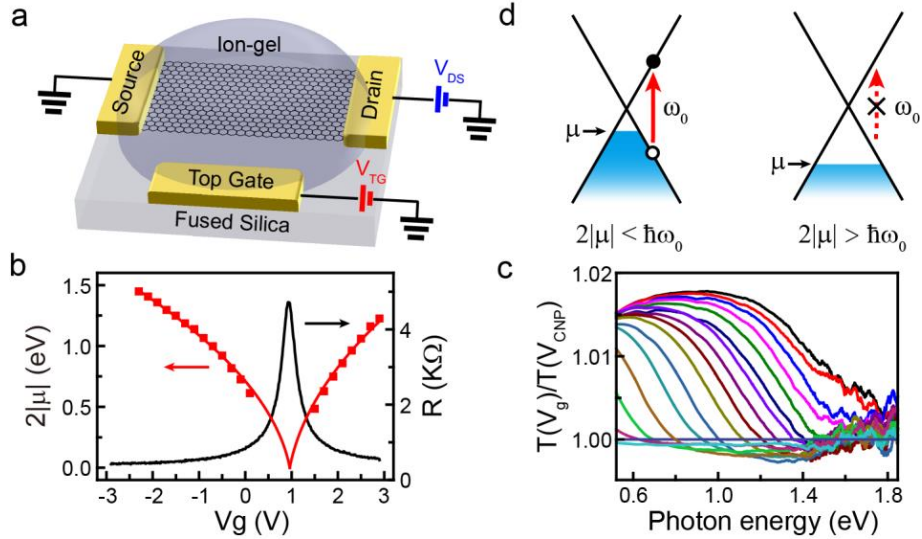

**Figure 1 | Tuning of chemical potential in graphene by ion-gel gating. a**, Schematic of an ion-gel gated graphene monolayer on a fused silica substrate covered by ion-gel and voltage-biased by the top gate. The source and drain electrodes on graphene are for resistance measurement. **b**, Measured graphene resistance as a function of gate voltage $V_g$ (black curve), the peak of which refers to the charge neutral point (CNP) or zero chemical potential ($\mu = 0$). The red squares and curve are $2|\mu|$ versus $V_g$ deduced from the transmittance spectra in **c** and calculated for the graphene device with an ion-gel capacitance of 2.5 μF/cm$^2$ (discussed in the SI), respectively. **c**, Transmittance spectra of graphene gated at different $V_g$–$V_{CNP}$, normalized against the one gated at $V_{CNP}$. Spectra from left to right correspond to $V_g$–$V_{CNP}$ changed from 0 to −3.2 V in steps of 0.2 V. The transmittance increases when $2|\mu| > \hbar\omega_0$. **d,** Linearly dispersed electronic bands of graphene around the CNP showing that tuning of $\mu$ enables Pauli blocking of interband transitions when $2|\mu| > \hbar\omega_0$.

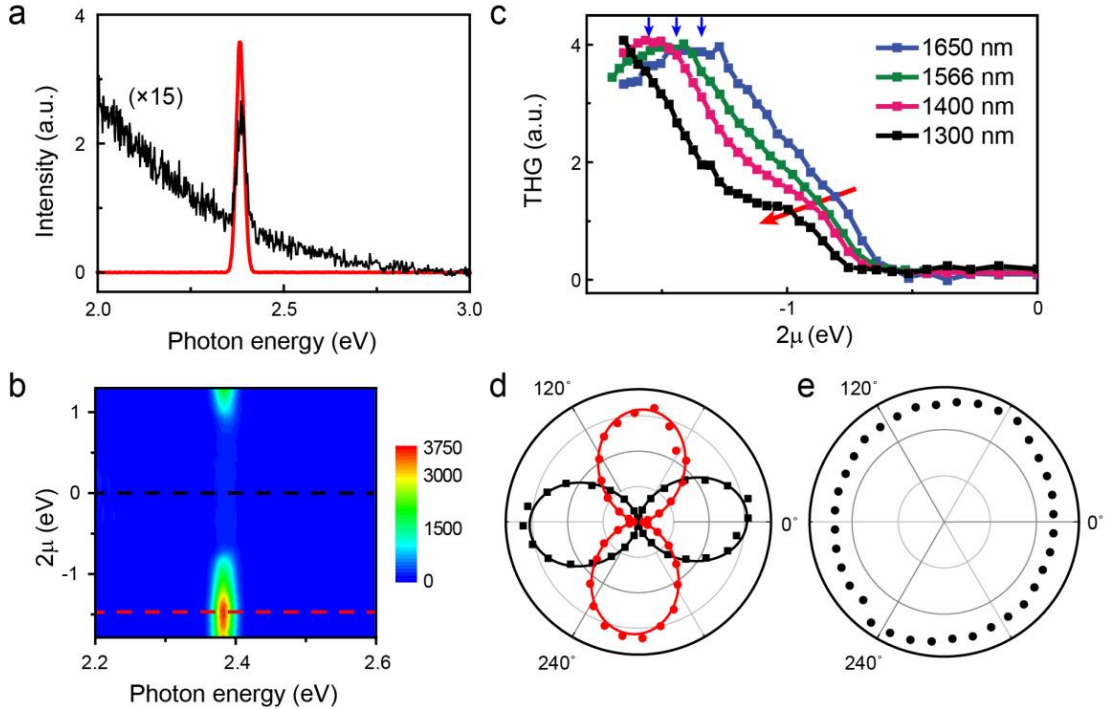

**Figure 2 | Gate-controlled THG from graphene and its polarization patterns. a**, Measured THG spectra by a normally incident femtosecond input pulse at 1566 nm from graphene gated at $\mu = 0$ (black curve, magnified by 15 times) and $\mu = -0.74$ eV (red curve), respectively. The broadband background of the black curve comes from up-converted photoluminescence due to rapid carrier-carrier scattering following one-photon interband excitation. The excitation power is 0.80 mW. **b**, Measured spectra versus $2\mu$ and photon energy showing strong dependence of THG at 2.38 eV with $\mu$. The spectra in **a** correspond to the signal variation following the black and red dashed lines. **c**, THG signal as a function of $2\mu$ generated by different input wavelengths: 1300 nm (black), 1400 nm (magenta), 1566 nm (green) and 1650 nm (blue). Curves are normalized for comparison. Dots are experimental data and curves are for eye guiding. Red and blue arrows mark the shoulder and maximum regions, respectively. **d** and **e**, Illustration that a linearly polarized input generates a linearly co-polarized THG output (with $\mu = -0.89$ eV). In **d**, THG output through an analyser is plotted as a function of angle $\theta$ between the analyser axis and the input polarization set along (black) and perpendicular to (red) the

source-drain directions. In both cases, the experimental data (dots) can be well fit by a $\cos^2\theta$ curve. In **e**, with the analyser axis parallel to the input polarization and rotating together azimuthally with respect to the sample, the THG output appears isotropic.

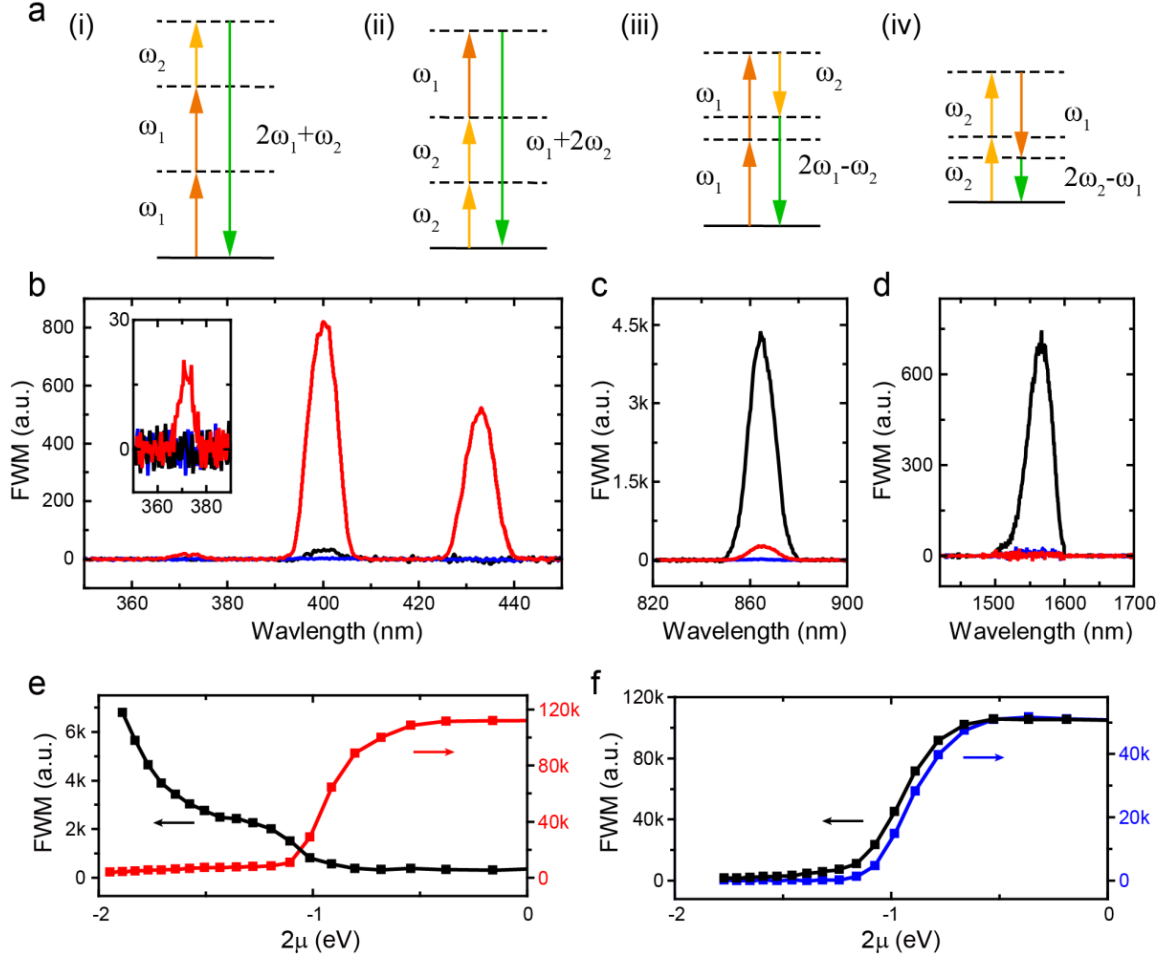

**Figure 3 | Four-wave mixing (FWM) in gated graphene. a,** FWM processes (i)-(iv) by two-colour excitation of $\omega_1$ and $\omega_2$ ($\omega_1 > \omega_2$). (i) and (ii) depict sum-frequency mixing (SFM) of $2\omega_1 + \omega_2$ and $\omega_1 + 2\omega_2$, respectively. (iii) and (iv) depict difference-frequency mixing (DFM) of $2\omega_1 - \omega_2$ and $2\omega_2 - \omega_1$, respectively. **b-d,** Output spectra of THG and SFM ($2\omega_1 + \omega_2$ at 371 nm, $\omega_1 + 2\omega_2$ at 400 nm, $3\omega_2$ at 433 nm), DFM ($2\omega_1 - \omega_2$ at 867 nm, and DFM ($2\omega_2 - \omega_1$ at 1566 nm), respectively. The input wavelengths are $\omega_1$ at 1040

nm and $\omega_2$ at 1300 nm for **b**, **c** and **e**, but $\omega_2$ is at 1250 nm for **d** and **f**. Black and red spectra spectra are for $\mu = 0$ and $\mu = -0.73$ eV, respectively. Blue spectra from the silica substrate are presented as reference. The inset in **b** shows the SFM signal at 371 nm, which is weak because of the limited sensitivity of our detector. **e,** Outputs of SFM ($\omega_1 + 2\omega_2$, black) and DFM ($2\omega_1 - \omega_2$, red) as functions of $2\mu$. Dots are experimental data and curves are for eye guiding. **f,** Outputs of DFM ($2\omega_1 - \omega_2$, black) and DFM ($2\omega_2 - \omega_1$, blue) as functions of $2\mu$.

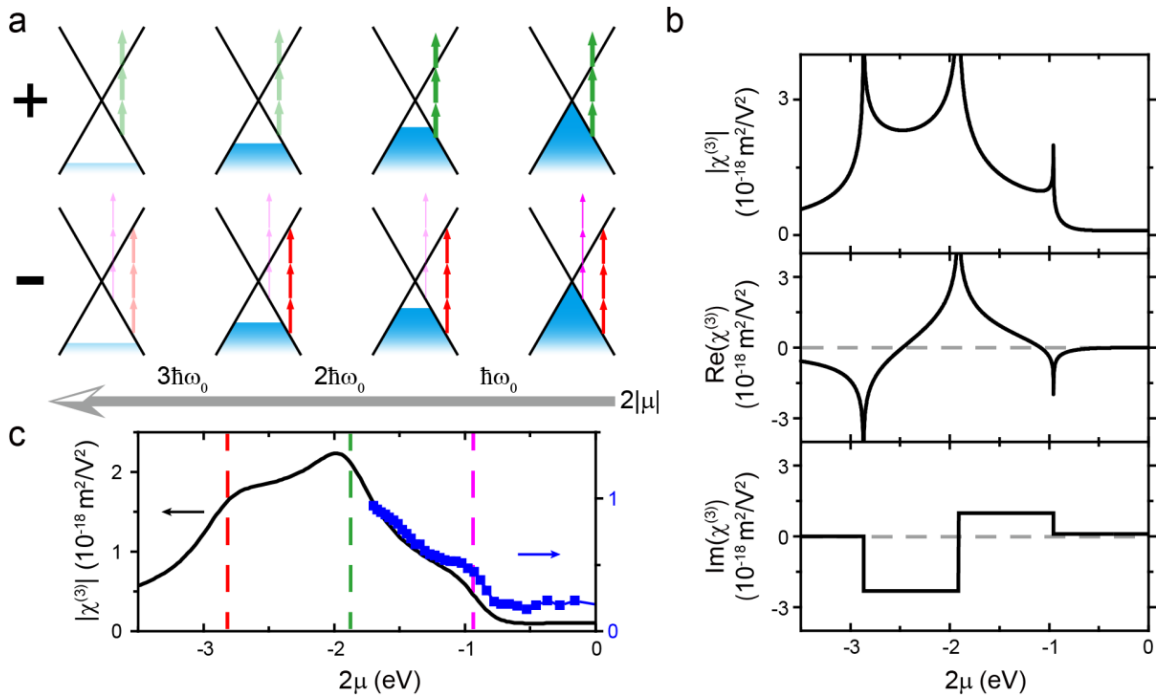

**Figure 4 | Theoretical understanding of $\mu$-dependent $\chi^{(3)}$ in THG. a,** Schematics showing how increase of $|\mu|$ successively switches off one-photon, two-photon, and three-photon interband transitions by Pauli-blocking in graphene. The switch-off is gradual at finite temperature and is described by the reduced brightness of the arrows. Two-photon transitions and one-, three-photon transitions contribute, respectively, to $\chi^{(3)}_{xxxx}$ positively and negatively. **b,** Calculated $\chi^{(3)}_{xxxx}$ versus $\mu$ for THG at $3\hbar\omega_0 = 2.868$ eV from graphene at zero temperature with resonant damping neglected, exhibiting

singularities at $|2\mu| = \hbar\omega_0$, $2\hbar\omega_0$, and $3\hbar\omega_0$. **c,** Comparison between experimental data (blue squares) and theoretical simulation (black curve) taking into account the finite temperature and resonant damping effects: T = 300 K, interband damping $\Gamma_e = 100\times|\mu|$ meV with $\mu$ in eV, and intraband damping $\Gamma_i = 0.5$ meV[29]. The dashed lines mark the positions of $|2\mu|=\hbar\omega_0$, $2\hbar\omega_0$ and $3\hbar\omega_0$.

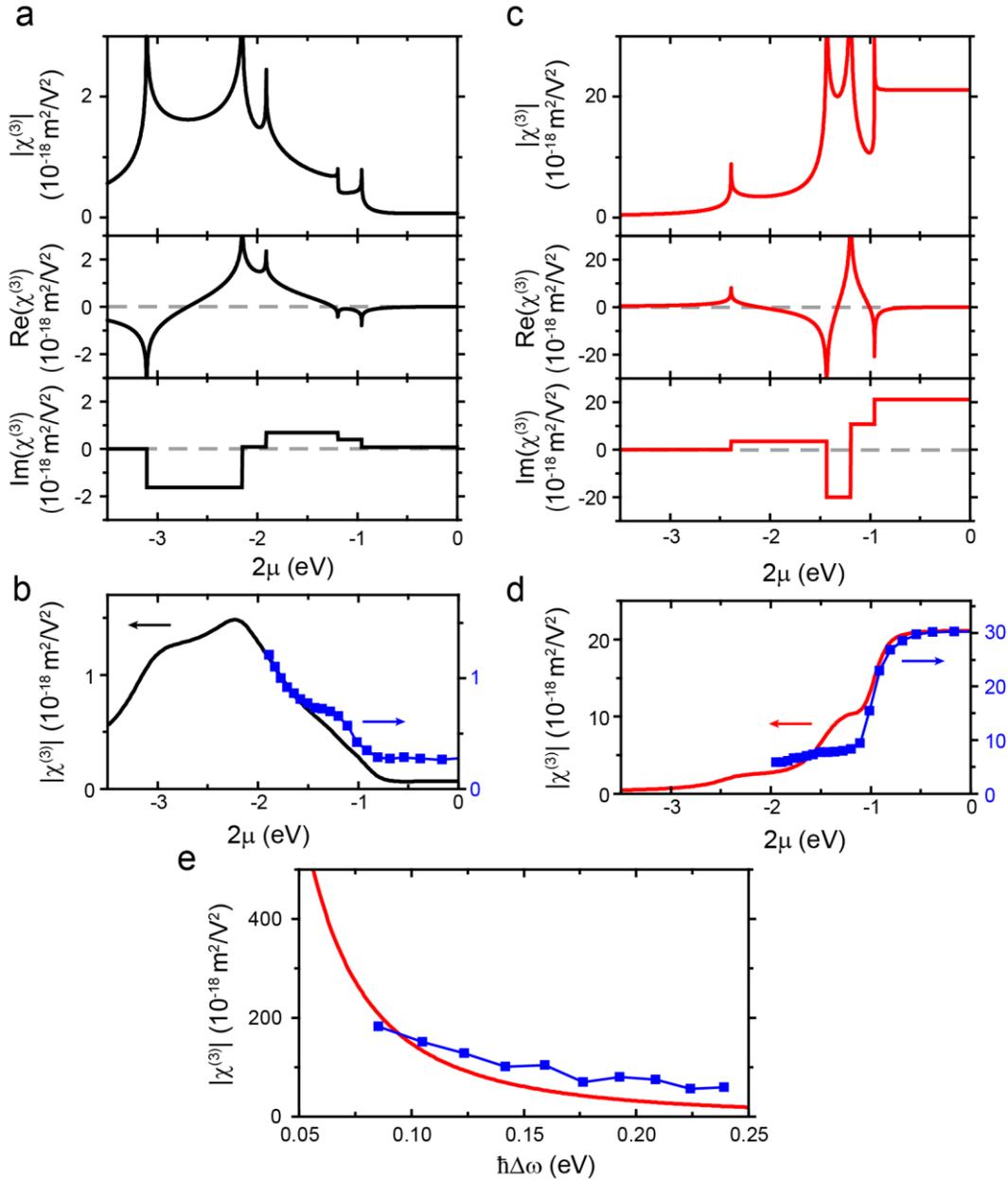

**Figure 5 | Theoretical calculations of $\mu$-dependent $\chi^{(3)}$ and comparison with experimental data for FWM from graphene**. **a, b,** Calculated $\chi^{(3)}_{xxxx}$ for $\omega_1 + 2\omega_2$ SFM at T = 0 K with no resonant damping and at finite temperature (300 K) with damping, respectively. **c, d,** Calculated $\chi^{(3)}_{xxxx}$ for $2\omega_1 - \omega_2$ DFM at T = 0 K with no resonant damping and at finite temperature (300 K) with damping, respectively. $\hbar\omega_1 = 1.195$ eV and $\hbar\omega_2 = 0.956$ eV. Corresponding experimental data (blue squares) are presented in **b** and **d** for comparison. **e,** Calculated $|\chi^{(3)}_{xxxx}|$ for undoped graphene ($\mu = 0$) as a function of $\Delta\omega$ (= $\omega_1 - \omega_2$) at T = 0 K with no damping showing divergence toward $\Delta\omega = 0$. Blue squares are experimental data for $2\omega_1 - \omega_2$ DFM with $\omega_1$ fixed at 1040 nm and $\omega_2$ tuned from 1120 nm to 1300 nm.